\RequirePackage{ifpdf}
\ifpdf 
\documentclass[pdftex]{sigma}
\else
\documentclass{sigma}
\fi

\numberwithin{equation}{section}

\begin{document}

\newcommand{\undertilde}[1]{\underset{\widetilde{}}{#1}}
\newcommand{\al}{\alpha}
\newcommand{\de}{\delta}
\newcommand{\De}{\Delta}
\newcommand{\be}{\beta}
\newcommand{\ga}{\gamma}
\newcommand{\Ga}{\Gamma}
\newcommand{\Be}{\Xi}
\newcommand{\Al}{\Lambda}

\newcommand{\ala}{\bar{\alpha}}
\newcommand{\alb}{\bar{\bar{\alpha}}}
\newcommand{\dea}{\bar{\delta}}
\newcommand{\bea}{\bar{\beta}}
\newcommand{\gaa}{\bar{\gamma}}

\newcommand{\als}{\hat{\alpha}}
\newcommand{\des}{\hat{\delta}}
\newcommand{\bes}{\hat{\beta}}
\newcommand{\gas}{\hat{\gamma}}

\newcommand{\as}{\hat{a}}
\newcommand{\ds}{\hat{d}}
\newcommand{\bs}{\hat{b}}
\newcommand{\cs}{\hat{c}}

\newcommand{\la}{\lambda}
\newcommand{\La}{\Lambda}
\newcommand{\laa}{\bar{\lambda}}
\newcommand{\lab}{\bar{\bar{\lambda}}}
\newcommand{\lac}{\bar{\bar{\bar{\lambda}}}}
\newcommand{\lad}{\overset{4}{\lambda}}
\newcommand{\lae}{\overset{5}{\lambda}}
\newcommand{\laf}{\overset{6}{\lambda}}

\newcommand{\mus}{\hat{\mu}}

\newcommand{\va}{\bar{v}}
\newcommand{\vb}{\bar{\bar{v}}}
\newcommand{\vs}{\hat{v}}
\newcommand{\vas}{\hat{\bar{v}}}

\newcommand{\ua}{\bar{u}}
\newcommand{\ub}{\bar{\bar{u}}}
\newcommand{\us}{\hat{u}}
\newcommand{\uas}{\hat{\bar{u}}}

\newcommand{\ta}{\bar{t}}
\newcommand{\tb}{\bar{\bar{t}}}
\newcommand{\ts}{\hat{t}}
\newcommand{\tas}{\hat{\bar{t}}}

\newcommand{\vo}{v_1}
\newcommand{\voa}{\bar{v}_1}
\newcommand{\vos}{\hat{v}_1}
\newcommand{\voas}{\hat{\bar{v}}_1}

\newcommand{\vt}{v_2}
\newcommand{\vta}{\bar{v}_2}
\newcommand{\vts}{\hat{v}_2}
\newcommand{\vtas}{\hat{\bar{v}}_2}

\newcommand{\vth}{v_3}
\newcommand{\vtha}{\bar{v}_3}
\newcommand{\vths}{\hat{v}_3}
\newcommand{\vthas}{\hat{\bar{v}}_3}

\newcommand{\vf}{v_4}
\newcommand{\vfa}{\bar{v}_4}
\newcommand{\vfs}{\hat{v}_4}
\newcommand{\vfas}{\hat{\bar{v}}_4}

\newcommand{\wa}{\bar{w}}
\newcommand{\ws}{\hat{w}}
\newcommand{\was}{\hat{\bar{w}}}
\newcommand{\wb}{\bar{\bar{w}}}
\newcommand{\wc}{\bar{\bar{\bar{w}}}}

\newcommand{\xa}{\bar{x}}
\newcommand{\xb}{\bar{\bar{x}}}
\newcommand{\xc}{\bar{\bar{\bar{x}}}}
\newcommand{\xd}{\overset{4}{x}}
\newcommand{\xe}{\overset{5}{x}}
\newcommand{\xf}{\overset{6}{x}}

\newcommand{\xs}{\hat{x}}
\newcommand{\xss}{\hat{\hat{x}}}
\newcommand{\xas}{\hat{\bar{x}}}
\newcommand{\xbs}{\hat{\bar{\bar{x}}}}

\newcommand{\ya}{\bar{y}}
\newcommand{\yb}{\bar{\bar{y}}}
\newcommand{\yc}{\bar{\bar{\bar{y}}}}
\newcommand{\yd}{\overset{4}{y}}
\newcommand{\ye}{\overset{5}{y}}
\newcommand{\yf}{\overset{6}{y}}

\newcommand{\ys}{\hat{y}}
\newcommand{\yss}{\hat{\hat{y}}}
\newcommand{\yas}{\hat{\bar{y}}}

\newcommand{\zs}{\hat{z}}
\newcommand{\za}{\bar{z}}
\newcommand{\zss}{\hat{\hat{z}}}
\newcommand{\zas}{\hat{\bar{z}}}

\newcommand{\ra}{\bar{r}}
\newcommand{\rb}{\bar{\bar{r}}}
\newcommand{\rc}{\bar{\bar{\bar{r}}}}
\newcommand{\rs}{\hat{r}}
\newcommand{\rss}{\hat{\hat{r}}}
\newcommand{\ras}{\hat{\bar{r}}}

\newcommand{\du}[3]{#1_{#2}^{#3}}
\newcommand{\ol}[1]{\overline{#1}}
\newcommand{\ul}[1]{\underline{#1}}
\newcommand{\os}[2]{\overset{#1}{#2}}

\newcommand\qon{{\rm qP}_{\rm{\scriptstyle I}}}
\newcommand\qtw{{\rm qP}_{\rm{\scriptstyle II}}}
\newcommand\qth{{\rm qP}_{\rm{\scriptstyle III}}}
\newcommand\qfi{{\rm qP}_{\rm{\scriptstyle V}}}
\newcommand\qsi{{\rm qP}_{\rm{\scriptstyle VI}}}

\newcommand\don{{\rm dP}_{\rm{\scriptstyle I}}}
\newcommand\dtw{{\rm dP}_{\rm{\scriptstyle II}}}
\newcommand\dth{{\rm dP}_{\rm{\scriptstyle III}}}

\newcommand\con{{\rm cP}_{\rm{\scriptstyle I}}}
\newcommand\ctw{{\rm cP}_{\rm{\scriptstyle II}}}

\allowdisplaybreaks

\renewcommand{\thefootnote}{$\star$}

\renewcommand{\PaperNumber}{089}

\FirstPageHeading

\ShortArticleName{A Completeness Study on Certain $2\times2$ Lax Pairs Including Zero Terms}

\ArticleName{A Completeness Study on Certain $\boldsymbol{2\times2}$ Lax Pairs\\
 Including Zero Terms\footnote{This
paper is a contribution to the Proceedings of the Conference ``Symmetries and Integrability of Dif\/ference Equations (SIDE-9)'' (June 14--18, 2010, Varna, Bulgaria). The full collection is available at \href{http://www.emis.de/journals/SIGMA/SIDE-9.html}{http://www.emis.de/journals/SIGMA/SIDE-9.html}}}

\Author{Mike C.~HAY}

\AuthorNameForHeading{M.C.~Hay}

\Address{Institute of Mathematics for Industry, Kyushu University, Ito Campus,\\
744 motooka, nishi-ku, Fukuoka, 819-0395, Japan}
\Email{\href{mailto:hay@math.kyushu-u.ac.jp}{hay@math.kyushu-u.ac.jp}}

\ArticleDates{Received April 09, 2011, in f\/inal form September 04, 2011;  Published online September 14, 2011}

\Abstract{We expand the completeness study instigated in [{\em J.~Math. Phys.}  {\bf 50}  (2009), 103516, 29~pages] which found all $2\times2$ Lax pairs with non-zero, separable terms in each entry of each Lax matrix, along with the most general nonlinear systems that can be associated with them. Here we allow some of the terms within the Lax matrices to be zero. We cover all possible Lax pairs of this type and f\/ind a new third order equation that can be reduced to special cases of the non-autonomous lattice KdV and lattice modif\/ied KdV equations among others.}


\Classification{37K15; 39A14; 70H06}

\section{Introduction}

Because this article expands on the notions introduced in \cite{h09}, we refer the reader to that paper for a more in depth introduction and problem description. The dif\/ference between the earlier work and this one is that in the earlier study all of the terms in each entry of the Lax matrices were assumed to be non-zero. Here we extend the analysis on $2\times2$ Lax pairs with a separable term in each entry of each matrix by allowing terms to be zero. Including zero terms covers some asymmetric Lax pairs that did not arise in the previous study, some of which are associated with quad graph equations that fall outside the Adler--Bobenko--Suris (ABS) list~\cite{abs03} because they do not possess the tetrahedron property, or are not even multidimensionally consistent.

Brief\/ly, the procedure used to analyze the Lax pairs runs as follows. The compatibility condition produces four expression, one for each entry of the $2\times 2$ Lax matrices, see~\eqref{cc} below. Each of these expressions can be split into smaller equations in various ways depending on how the spectral parameter is chosen. Thus, the compatibility condition def\/ines various systems of equations that we subsequently solve, in a manner that preserves their full generality, up to a~point where a nonlinear evolution equation is apparent, or it has been shown that the system cannot be associated with a nonlinear system. Testing all combinations of terms, we thereby survey the complete set of Lax pairs of the type described.

The paper is organized as follows. Section~\ref{sec:results} summarizes the chief results including a new third order system and a theorem covering the relevant Lax pairs. Section~\ref{proof} describes the method used to identify and analyze all of the Lax pairs considered. Section~\ref{sec:rel} places the results found here in the context of those in the literature. And a discussion rounds out the paper.

\section{Results}\label{sec:results}

There are two principal results reported in this paper. The f\/irst is an apparently new partial dif\/ference equation which can be conveniently written as a coupled pair as follows
\begin{gather}\label{main}
\frac{\xas}{\xs}+y = \frac{\xas}{\xa},\qquad 
\frac{\xa}{x}+\ya = \frac{\xs}{x},
\end{gather}
where $x(l,m)$ and $y(l,m)$ are the dependent variables over the discrete independent variables $l$ and $m$,  shifts in which are represented by $\bar{\cdot}$ and $\hat{ \cdot}$ respectively.
One may easily isolate $y$ from either equation and use it to form a single equation in~$x$ which is second order in~$l$ and f\/irst order in~$m$. The derivation of~\eqref{main} is given in Section~\ref{sec:l22}. We prefer to write the system as a pair of equations because from this form it is easier to see how it can be reduced to scalar equations on quad graphs, see Section~\ref{sec:rel}. Note that no parameters or arbitrary functions, which are usually associated with the lattice spacing, are present in~\eqref{main} because they can all be removed by a~gauge. However, they do appear in the reduced equations on quad graphs in Section~\ref{sec:rel}.

A Lax pair for \eqref{main} is
\begin{gather}
L=\begin{pmatrix} \xa/x&\nu \\\nu&0\end{pmatrix},\qquad  M=\begin{pmatrix}\xs/x&\nu \\ \nu&y\end{pmatrix},\label{mainLP}
\end{gather}
where $\nu$ is the spectral parameter.

Also relevant to this article are the higher order analogues of the lattice modif\/ied KdV equation (LMKdV) and lattice sine-Gordon equation (LSG), called LMKdV$_2$ and LSG$_2$, found in~\cite{h09}.

This article's second principal result is the following theorem.

\begin{theorem} \label{theorem} The system of equations that arise via the compatibility condition of any $2\times2$ Lax pair with separable terms in each entry of each matrix $($as described in {\rm \cite{h09})} is either trivial, underdetermined, overdetermined, or can be reduced to one of {\rm LMKdV$_2$} or {\rm LSG$_2$}, a degenerate form of those equations, or a transformation from one of those.
\end{theorem}

The transformations we refer to in Theorem~\ref{theorem} are M\"obius, Miura or B\"acklund transformations, which allow one to use the same Lax pair for the new equation, after the change of variables has been applied to it as well.

For the cases when none of the entries in the Lax matrices are zero, the proof of Theorem~\ref{theorem} was carried out in~\cite{h09} and depends on considering all of the possible sets of equations that can arise from the compatibility condition of such Lax pairs. Section~\ref{proof} of this paper details the proof for the cases when at least one term in at least one of the Lax matrices takes the value zero.

\section{Analysis of all relevant Lax pairs}\label{proof}
This paper relies heavily on the methodology reported in \cite{h09}, see Section~2 of that paper for a~detailed  description of the techniques used here.

As the compatibility condition (CC) is central to the entire thread of this paper, we write it out here and refer back to it throughout. Taking the general form of the Lax matrices to be
\begin{gather*}
L=\begin{pmatrix} aA&bB\\cC&dD\end{pmatrix},\qquad  M=\begin{pmatrix}\al\Al&\be\Be\\\ga\Ga&\de\De\end{pmatrix},
\end{gather*}
where, as throughout this paper, lower case letters depend on the lattice variables only, $a=a(l,m)$, and upper case letters depend on the spectral variable only, $A=A(\nu)$.

Separating the CC, $\widehat{L}M=\ol{M}L$, into each of its entries gives:
\begin{subequations}\label{cc}
\begin{alignat}{3}
& (1,1) \quad && \as \al A\Al+\bs\ga B\Ga = a\ala A\Al+c\bea C\Be,& \label{CC11}\\
& (1,2)\quad && \as \be A\Be+\bs\de B\De = b\ala B\Al+d\bea D\Be, & \label{CC12}\\
& (2,1)\quad && \ds \ga D\Ga+\cs\al C\Al = c\dea C\De+a\gaa A\Ga, & \label{CC21}\\
& (2,2)\quad && \ds\de D\De+\cs\be C\Be = d\dea D\De+b\gaa B\Ga. & \label{CC22}
\end{alignat}
\end{subequations}

To solve the systems of equations that arise via the CC of each Lax pair, we use the method described in \cite[Section~III]{h09}.

We will examine Lax pairs on a case by case basis based on the number of their entries which are zero. The following proposition is therefore useful.
\begin{proposition}\label{prop00} If there are two or more zero entries in either Lax matrix then the resulting compatibility condition is either linear or underdetermined.
\end{proposition}

\begin{proof}
Let the required two zero terms reside in the matrix $L$. Then there are three cases that need to be considered.

{\bf Case 1.}  Diagonal entries of~$L$ are zero. Applying the compatibility condition to a Lax pair with zero diagonals is achieved by setting $A=D=0$ in~\eqref{cc}. This shows that all the resulting conditions are linear. Thus, this Lax pair cannot be associated with a nonlinear system, unless the nonlinearity is arbitrarily introduced into a Lax pair that possesses excess freedom. We call such Lax pairs with excess freedom `false' (see the argument in Appendix~\ref{app:10only}, case~4, or Section~4.3 of~\cite{h09}). We remark that two linear equations may result in a nonlinear equation so long as one is additively linear and the other is multiplicative. An example of this can be seen in~\cite[equation~(7)]{gpr97}. However, in this case, all equations are linear in a multiplicative sense.

{\bf Case 2.}  Of\/f diagonal entries are zero. Like the previous case, this also leads to linear conditions only.

{\bf Case 3.}  One row or column is zero. Let the right column of $L$ be zero,   i.e.\ set $B$ and $D$ to zero in~\eqref{cc}. Then the $(1,2)$ entry of~\eqref{cc} is $a\be A\Be=0$, which we solve by taking $\Be=0$ because allowing $A=0$ would cause the remaining conditions to all be linear. Now the $(1,2)$ and $(2,2)$ entries are both identities while the~$(1,1)$ entry gives a condition that can be solved by introducing the new dependent variable~$v(l,m)$ as follows
\begin{gather*}
(1,1)\quad\as \al=a\ala\quad\Rightarrow\quad a=\la \va/v, \quad\al=\mu\vs/v,
\end{gather*}
where, as throughout this paper, $\la=\la(l)$ and $\mu=\mu(m)$ are arbitrary functions. Finally, consider the $(2,1)$ entry of the compatibility condition
\begin{gather*}
\cs\al C\Al=a\gaa A\Ga+c\dea C\De.
\end{gather*}
We are able to choose the spectral terms such that all of the lattice terms remain together in one equation, or such that they are separated into more than one equation. The former case yields the sole condition
\begin{gather*}
\frac{\cs\al}{c\dea}=\frac{a\gaa}{c\dea}+1,
\end{gather*}
but freedom remains here to the extent that any equation can be written into this form by choosing the remaining lattice terms appropriately. Thus, this is a false Lax pair. The latter case yields more equations, which may appear to solve the problem of having excess freedom, however, the resulting equations must all be linear, resulting in a trivial system.
\end{proof}

We are now in a position to analyze all of the possible Lax pairs of the category described in the introduction. Since the more interesting results occur when there are less zero entries in the Lax matrices, we start with just a single zero in one matrix and work our way toward the simpler systems.

\subsection{A single zero term in one Lax matrix only}
We assume that the zero lies in the $L$ matrix. There are only two cases to consider, one with the zero in a diagonal entry and one in an of\/f-diagonal entry.

\subsubsection{A zero in a diagonal entry}\label{sec:l22}

$L=\begin{pmatrix}aA&bB\\cC&0\end{pmatrix}$, $M=\begin{pmatrix}\al\Al&\be\Be\\\ga\Ga&\de\De\end{pmatrix}$. Setting $D=0$ in~\eqref{cc} shows that the spectral terms products in the~(2,2) entry must be equal,  i.e.~$B\Ga=C\Be$. This can be seen by making considerations similar to those in~\cite[Section~II]{h09}. As such, there are four cases that require further consideration, these are listed in Table~\ref{tab:l22}. The analysis is similar for all four cases listed but only the f\/irst yields a major result, so cases~2,~3 and~4 are dealt with in Appendix~\ref{app:10only}, while case~1 is described below.

\begin{table}[h!]
\centering
\caption{The four cases that require consideration when $D=0$ in \eqref{cc}. The lines represent proportionality between the spectral term products. In cases~3 and~4, $A\Be$ and $A\Ga$, respectively, are split into two terms not proportional to one another.}\label{tab:l22}
\vspace{1mm}

\begin{tabular}{|c| c|c |c| c|}
\hline
Case&$\begin{array}{c} (1,1) \\
A\Al\quad B\Ga\\\quad \end{array}$ &$\begin{array}{c} (1,2)\\
\mbox{\footnotesize{$A\Be$}}\begin{array}{c}\mbox{\footnotesize{$B\Al$}}\\
\mbox{\footnotesize{$B\De$}}\end{array}\end{array}$&$\begin{array}{c} (2,1)\\
\mbox{\footnotesize{$A\Ga$}}\begin{array}{c}\mbox{\footnotesize{$C\Al$}}\\
\mbox{\footnotesize{$C\De$}}\end{array}\end{array}$\tsep{14pt} &Result\bsep{2pt}\\
\hline 1& not $\propto$
&\includegraphics{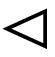}
&\includegraphics{Hay-Fig1}
&new nontrivial system\tsep{8pt}\\
  2&$\propto$&\includegraphics{Hay-Fig1}
  &\includegraphics{Hay-Fig1}
  &underdetermined\tsep{8pt}\\
 3&not $\propto$&\includegraphics{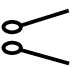}
 &\includegraphics{Hay-Fig2}
 &false Lax pair\tsep{8pt}\\
 4&$\propto$&\includegraphics{Hay-Fig2}
 &\includegraphics{Hay-Fig2}
 &false Lax pair\tsep{8pt}\bsep{2pt}\\
\hline
\end{tabular}

\end{table}

{\bf Case 1.} Starting with the CC \eqref{cc}, setting $D=0$ and separating the (1,1) entry into two equations yields the following
\begin{alignat}{5}
& (1,1)_1 \quad & &\as\al=a\ala,\qquad &&(1,2)\quad &&\as\be+\bs\de=b\ala,& \nonumber\\
& (1,1)_2 \quad && \bs\ga=c\bea, && &&&\nonumber \\
& (2,1)\quad&& \cs\al=a\gaa+c\dea,\qquad &&(2,2)\quad&&\cs\be=b\gaa.&
 \label{l22c1cc}
\end{alignat}
Solving the diagonal entries f\/irst, which are linear in this case, we f\/ind that (1,1)$_1$ leads to
\begin{gather}
a=\la_1\va/v,\qquad \al=\mu_1\vs/v,\label{l22c1values1}
\end{gather}
where we have introduced the new unknown $v(l,m)$ and the arbitrary functions $\la_1(l)$ and $\mu_1(m)$. Equations (1,1)$_2$ and $(2,2)$ from \eqref{l22c1cc} are solved in tandem leading to
\begin{gather}
b=\la_2\rho\frac{\ta}{u},\qquad c=\frac{\la_2}{\rho}\frac{\ua}{t},\qquad \be=\frac{\mu_2}{\sigma}\frac{\ts}{u},\qquad \ga=\mu_2\sigma\frac{\us}{t},\label{l22c1values2}
\end{gather}
where we have introduced the new unknowns $t(l,m)$ and $u(l,m)$, as well as the arbitrary functions $\la_2(l)$, $\mu_2(m)$, $\rho=\la_3(l)^{(-1)^m}$ and $\sigma=\mu_3(m)^{(-1)^l}$. Substituting \eqref{l22c1values1} and \eqref{l22c1values2} into the of\/f diagonal entries of \eqref{l22c1cc} shows that certain combinations of variables are repeated allowing us to set~$t\equiv u\equiv 1$, w.l.o.g. We also rename $v=x$, $\de=y$, and use gauge transformations to remove all arbitrary functions. This leads to~\eqref{main}.

\subsubsection{A zero in an of\/f-diagonal entry}
$L=\begin{pmatrix}aA&0\\cC&dD\end{pmatrix}$, $M=\begin{pmatrix}\al\Al&\be\Be\\\ga\Ga&\de\De\end{pmatrix}$. The equations from the CC are obtained by setting $B=0$ in \eqref{cc}, these are
\begin{alignat*}{5}
& (1,1)\quad && (\as\al-a\ala) A\Al=c\bea C\Be,\qquad &&(1,2)\quad&& \as\be A\Be=d\bea D\Be,& \\
& (2,1)\quad && \ds\ga D\Ga+\cs\al C\Al=a\gaa A\Ga+c\dea C\De,\qquad &&(2,2)\quad && (\ds\de-d\dea)D\De+\cs\be C\Be=0.&
\end{alignat*}
Since any extra zero entries in $L$ is covered by Proposition~\ref{prop00}, and in~$M$ will bring us to one of the Lax pairs considered in the other sections below (or Proposition~\ref{prop00}), we can make the following observations regarding the spectral terms. From the~(1,2) entry we have $A=D$. From the diagonal entries we have $A\Al=C\Be=D\De$ which implies that $\Al=\De$. Turning to the~(2,1) entry last, we f\/ind two pairs of repeated spectral terms $A\Ga$ and $C\Al$. These can either be chosen to be proportional to one another, so that only one lattice term equation arises in this entry, or separated to construct two equations.

Among the lattice terms there are seven unknowns, two of which are redundant after gauge transformations,  suggesting that f\/ive is the appropriate number of equations, and that the (2,1) entry should be separated into two equations. However, we must check both cases. Let us f\/irst examine the case where the (2,1) entry is split up, the case where the (2,1) entry is left as a~single equation is considered at the end of this subsection.

Dealing with the case where $A\Ga$ is not proportional to $C\De$ f\/irst, we have the following set of lattice term equations:
\begin{alignat*}{5}
& (1,1)\quad && \as\al-a\ala=c\bea,\qquad &&(1,2)\quad && \as\be =d\bea,& \\
& (2,1)_1\quad && \ds\ga =a\gaa,\qquad &&(2,2)\quad && \ds\de-d\dea+\cs\be =0,& \\
& (2,1)_2\quad && \cs\al=c\dea.&& &&&
\end{alignat*}
The three equations in the of\/f-diagonal entries are linear and have a general solution expressible as
\begin{gather*}
a=\frac{\la}{\rho}\frac{\ua}{u},\qquad d=\la\rho\frac{\va}{v},\qquad\al=\dea\frac{c}{\cs},\qquad\be=\mu_1\frac{\us}{v},\qquad\ga=\mu_2\frac{\vs}{u},
\end{gather*}
where all of the quantities are similar to those described below~\eqref{l22c1values2}. However, when these values are substituted into the remaining equations coming from the diagonal entries, we f\/ind that $c$ and $\de$ are redundant, so we set these to unity. Once again the arbitrary functions are~removable by gauge transformations, so we rename $v=x$, $u=z$ and the f\/inal two equations are
\begin{subequations}\label{l12main}
\begin{gather}
\frac{\xa}{\xas}-\frac{x}{\xs} = \frac{\zs}{\xas},\label{l12main1}\\
\frac{z}{\zs}-\frac{\za}{\zas} = \frac{z}{\xa},\label{l12main2}
\end{gather}
\end{subequations}
where the dependent variables, $x$ and $z$, may be likened to $x$ and $y$, respectively, in equation~(1.2) from~\cite{h09}, which shows that this is a degenerate form of LSG$_2$. Obviously, one may isolate  either~$x$ or~$z$ from the appropriate equation in~\eqref{l12main}, substitute the result into the other equation and obtain an expression for one variable alone.

In fact, \eqref{l12main} is linearizable as follows.
Use \eqref{l12main1}  in \eqref{l12main2}  to obtain
\begin{gather}
\frac{\zs}{\hat{\zs}}=\frac{\xas}{\hat{\xas}}\cdot\frac{\xa/\xas-x/\xs}{\xas/\hat{\xas}-\xs/\hat{\xs}}.\label{l12p2}
\end{gather}
If we let $\psi=x/\xs$ then \eqref{l12p2} can be integrated twice for
\begin{gather*}
\hat{\psi}\psi-\mu_1\hat{\psi}=\mu_2,
\end{gather*}
which is a Riccati equation for $\psi$, where $\mu_{1,2}(m)$ are arbitrary functions, one of which can be removed by a gauge.

Having dealt with the case where $A\Ga$ is not proportional to $C\Al$, we must turn our attention to the case where these terms are proportional to one another, and the (2,1) entry of the compatibility condition is left as one equation (see the f\/irst paragraph of this subsection). This case has too few conditions relative to the number of unknowns and is expected to yield an underdetermined or false Lax pair, as is indeed the case. Similar analysis to the above shows that this Lax pair is false.

\subsection{A single zero term in each Lax matrix}
Due to symmetry considerations, only f\/ive Lax pairs with exactly one zero in each matrix need to be checked, these are listed in Table~\ref{tab10each}. All other Lax pairs of this type, with zeros located in dif\/ferent entries to those listed in Table~\ref{tab10each}, are equivalent to one of the Lax pairs listed. Of the f\/ive cases listed in Table~\ref{tab10each}, only the f\/irst will be explained here. The analysis of the other four cases is similar to the f\/irst case so these are dealt with in Appendix~\ref{app:10each}.
\begin{table}[h!]\centering
\caption{Lax pairs with one zero in each matrix.}\label{tab10each}

\vspace{1mm}

\begin{tabular}{|c |c|}
\hline
 $L$&$M$ \tsep{2pt}\bsep{2pt}\\
\hline
$\begin{pmatrix}aA&0\\cC&dD\end{pmatrix} $&$\begin{pmatrix}\al\Al&\be\Be\\0&\de\De\end{pmatrix}$, $\begin{pmatrix}\al\Al&0\\\ga\Ga&\de\De\end{pmatrix}$\tsep{10pt}\bsep{10pt}\\
\hline
$\begin{pmatrix} aA&bB\\cC&0\end{pmatrix}$&$\begin{pmatrix}\al\Al&\be\Be\\\ga\Ga&0\end{pmatrix}$, $\begin{pmatrix} \al\Al&0\\\ga\Ga&\de\De\end{pmatrix}$, $\begin{pmatrix} 0&\be\Be\\\ga\Ga&\de\De\end{pmatrix}$
\tsep{10pt}\bsep{10pt}\\
\hline
\end{tabular}
\end{table}


$L=\begin{pmatrix} aA&0\\cC&dD\end{pmatrix}$, $M=\begin{pmatrix}\al\Al&\be\Be\\0&\de\De\end{pmatrix}$. Setting $B=\Ga=0$ in~\eqref{cc} shows that, in order to prevent there being two zeros in one of the Lax matrices, the situation dealt with by Proposition~\ref{prop00}, we must have the following relationships between the spectral term products
\begin{gather*}
A\Al=C\Be,\qquad  D\De=C\Be,\qquad  A\Be=D\Be, \qquad  C\Al=C\De.
\end{gather*}
This combination of spectral terms brings about the following set of lattice term equations from the CC
\begin{alignat}{5}
& (1,1)\quad && \as\al-a\ala=c\bea,\qquad &&(1,2)\quad && \as\be=d\bea, & \nonumber\\
& (2,2)\quad && d\dea-\ds\de=\cs\be,\qquad &&(2,1)\quad && \cs\al=c\dea. &\label{l12m21cc}
\end{alignat}
From the (1,2) and (2,1) entries we take
\begin{gather}
d=\as\frac{\be}{\bea},\quad\quad\quad\al=\dea\frac{c}{\cs},\nonumber
\end{gather}
respectively. Substituting these values into the (1,1) and (2,2) entries of \eqref{l12m21cc}, we get two equations that can be rearranged to form
\begin{gather}\label{l12m21last2}
\as\dea-\frac{a\bar{\dea}\bar{c}\cs}{c\hat{\bar{c}}}=\cs\bea,\qquad
\frac{\hat{\as}\de\bes\bea}{\be\hat{\bea}}-\as\dea=-\cs\bea.
\end{gather}
These are added to give a linear equation that has the general solution
\begin{gather}
a=\la_1\rho\frac{\ta}{t},\qquad  c=\mu_2\frac{v}{\ts t},\qquad \de=\mu_1\sigma\frac{\us}{u},\qquad\be=\la_2\frac{v}{\ua u},\label{l12m21values}
\end{gather}
where we have introduced the dependent variables $t(l,m)$, $u(l,m)$, $v(l,m)$, as well as the arbitrary functions $\rho=\la_3^{(-1)^m}$ and $\sigma=\mu_3^{(-1)^l}$, where $\la_i=\la_i(l)$ and $\mu_i=\mu_i(m)$. Naturally, the values in~\eqref{l12m21values} cause both of equations~\eqref{l12m21last2} to be equivalent. In this f\/inal equation, we f\/ind that certain groups of lattice terms appear in the same combinations repeatedly, which is expected because of the excess freedom left in the Lax matrices that could be removed by gauge transformations. As such, we can remove $t$ and $u$ by setting, w.l.o.g.,
\begin{gather*}
t\equiv 1,\qquad u\equiv 1,\qquad v=1/x.
\end{gather*}
Noting that, in this case, all of the arbitrary functions can be removed by a gauge transformation, we f\/ind that either of the equations~\eqref{l12m21last2} becomes
\begin{gather}
\xa\xs-x\xas=1.\label{l12m21}
\end{gather}
As is shown in the next section, see \eqref{l12lsg} below, this is a degenerate form of LSG.

\section{Relationship with known results}\label{sec:rel}
In this section we describe the relationship between \eqref{main}, \eqref{l12main}, \eqref{l12m21} and some known systems.


It is clear that that setting $y=\mu x/\xs$ in \eqref{main} causes both equations to coincide and leads to
\begin{gather}
\xas(\xs-\xa)=\mu \xa x,\label{l22lmkdv}
\end{gather}
where the arbitrary function $\mu(m)$ cannot be removed by a gauge transformation. Since LMKdV can be written
\begin{gather*}
\xas(\xs-\xa)=x(\mu \xa-\la \xs),
\end{gather*}
equation \eqref{l22lmkdv} is a degenerate form of LMKdV with $\la=0$. We remark that the equation referred to here as LMKdV is also known as the lattice potential modif\/ied KdV equation, due to its continuum limit.

It is possible to transform \eqref{l22lmkdv} to a special case of the non-autonomous lattice KdV equation (LKdV),
\begin{gather}
\mu\ws+\frac{1}{\was}=\underset{\widehat{}}{\mu}\wa+\frac{1}{w},\label{l22lkdv}
\end{gather}
where $\underset{\widehat{}}{\mu}=\mu(m-1)$, which was introduced in~\cite{m07}. A scalar Lax pair for the non-autonomous LKdV was found in~\cite{ko08}, while the autonomous form and associated scalar Lax pair were f\/irst given in~\cite{h77}, the autonomous form was also shown to pass the singularity conf\/inement test~\cite{ghrv09}. The Miura transformation
\begin{gather*}
\frac{\xa}{x}=\underset{\widehat{}}{\mu}\wa+\frac{1}{w},\qquad\frac{\xs}{x}=\mu \ws,
\end{gather*}
transforms \eqref{l22lmkdv} to \eqref{l22lkdv}.

The following equation was introduced in an autonomous form in \cite{nc95}
\begin{gather}
\bea(\mu\be+1)=\bes(\mus\hat{\bar{\beta}}+1),\label{levi}
\end{gather}
where it appears as equation~(6.15). In a dif\/ferent, non-autonomous form, equation~\eqref{levi} also appears as~(31) in~\cite{ly09}, where a Miura transformation between it and the Volterra equation was given, as well as a $2\times2$ Lax pair, a result which drew comment in~\cite{rgsw09}. Since the Lax pair given in~\cite{ly09} is equivalent to the one for~\eqref{main} in terms of the spectral dependence, see~\eqref{mainLP}, we expect there to exist a relationship between~\eqref{main} and~\eqref{levi}. Indeed, after reducing~\eqref{main} to~\eqref{l22lmkdv} as shown, the following Miura transformation relates it to \eqref{levi}
\begin{gather*}
\frac{x}{\xas}=\be,\qquad\frac{\xs}{\xa}=\mu\be+1.
\end{gather*}

Note that neither \eqref{l22lkdv} nor \eqref{levi} has been found to be multidimensionally consistent and they are consequently left undetected by studies such as \cite{abs03,h04,h05,b10,as04} that rely on this property.

Of course, Lax pairs for any of these equations can be written down by making the appropriate substitutions in~\eqref{mainLP}. It is a simple matter to put the $2\times 2$ Lax pair for LKdV found here into scalar form so that it can be compared to previous results from~\cite{ko08,h77}. On doing so one can see that it is equivalent to the scalar Lax pair found in~\cite{h77}, except that the appearance of~$\mu$ makes~\eqref{l22lkdv} an extension of that autonomous case, while it is not equivalent to the full non-autonomous Lax pair found in~\cite{ko08}. We believe that~\eqref{main} is the most general system that can be associated with the Lax pair found in~\cite{h77}, and we expect that the full non-autonomous form of LKdV is a reduction of some higher order system that can be associated with its more complicated Lax pair, which contains more terms.

Although \eqref{l12main} was shown to be linearizable, it is still interesting to describe its place among known equations. One may reduce~\eqref{l12main} to a scalar equation on a quad-graph as was done in the reduction of \eqref{main} to the scalar equation~\eqref{l22lmkdv}. While we may introduce an arbitrary function in the reduction to~\eqref{l12lsg}, $z\rightarrow \la\mu/x$ say, in this case the arbitrary function is removable by a~gauge. As such, we make the change of variables $z\rightarrow1/ x$, causing~\eqref{l12main} to become
\begin{gather}\label{l12lsg}
\xa\xs-x\xas=1,
\end{gather}
which can be considered a degeneration of LSG or an extension of equation 1.2n in \cite{h05}, since in that paper the right hand side was zero. Equation \eqref{l12lsg} has not been found to be multidimensionally consistent and it does not possess the tetrahedron property. To see that this is a~degeneration of LSG, we put that equation into the form
\begin{gather*}
\xa\xs-x\xas=1-\la\mu x\xa\xs\xas,
\end{gather*}
which is evidently equivalent to \eqref{l12lsg} if we choose either of the arbitrary functions $\la(l)$ or $\mu(m)$ to be zero.

As expected since \eqref{l12main} is linearizable, equation \eqref{l12lsg} is also linearizable in the same way.

\section{Discussion}\label{disc}

The present study compliments the novel and rich results provided by the multidimensional consistency approach (MCA). MCA produces many integrable equations on quad graphs, all of which have a readily derivable Lax pair. However, as the present study shows, direct analysis of the Lax pairs can lead to equations that appear not to be multidimensionally consistent. The tetrahedron property is a common assumption in MCA studies, although some work has been done to f\/ind equations that do not possess it. The present study also f\/inds a system that does not possess the tetrahedron property, however, like many of the other known results without it~\mbox{\cite{h05,rjgt06}}, this system is linearizable.

It is interesting to relate the results found here to those of \cite{h09}. There the following  coupled pair of nonlinear equations was found
\begin{subequations}\label{lmkdv2a}
\begin{alignat}{3}
& \text{{\rm LMKdV$_2$:}}\quad && \frac{\xas}{\xs}+\frac{\zs}{z} =   \la_1\frac{\za}{z}+\frac{\xas}{\xa},& \\
&&& \frac{\xs}{x} +\la_1\frac{\zas}{\zs} = \frac{\xa}{x}+\frac{\zas}{\za},&
\end{alignat}
\end{subequations}
which is LMKdV$_2$ with all but one arbitrary function, $\la_1(l)$, removed by a gauge. LMKdV$_2$ becomes LMKdV when $z=\la\mu/x$, and can also be transformed to a two-component vector analogue of the cross ratio equation, also known as the discrete Schwarzian KdV equation. However, unlike the degenerate form of LMKdV \eqref{l22lmkdv},  no such transformation to LKdV from the full LMKdV is evident. A Lax pair for \eqref{lmkdv2a} is
\begin{gather}
L=\begin{pmatrix} \xa/x&\nu\\ \nu&\la_1\za/z\end{pmatrix},\qquad M=\begin{pmatrix} \xs/x&\nu\\\nu&\zs/z\end{pmatrix}.\label{lmkdv2lp}
\end{gather}
Compare this to the Lax pair found in the present study~\eqref{mainLP} whose compatibility condition is~\eqref{main}, where we could let $y=\zs/z$ and f\/ind
\begin{gather}\label{main3}
\frac{\xas}{\xs}+\frac{\zs}{z} = \frac{\xas}{\xa},\qquad 
\frac{\xa}{x}+\frac{\zas}{\za} = \frac{\xs}{x}.
\end{gather}
It is now obvious that allowing the (2,2) entry of $L$ in~\eqref{lmkdv2lp} to be zero by taking $\la_1=0$, so that we get~\eqref{mainLP}, simply removes the corresponding terms from~\eqref{lmkdv2a} and produces~\eqref{main3}. Thus, the special case of LKdV~\eqref{l22lkdv}, along with~\eqref{levi}, can, in some sense, be thought of as degenerations of LMKdV$_2$. This is suggestive of an underlying structure between various discrete integrable systems which is related to their Lax pairs. However we withhold further comment on the matter until future studies are completed that include more terms in the Lax matrices.

\appendix

\section[Remaining cases with one zero in $L$ only]{Remaining cases with one zero in $\boldsymbol{L}$ only}\label{app:10only}

This appendix deals with the cases 2, 3 and 4 from Table~\ref{tab:l22} in Section~\ref{sec:l22}.

{\bf Case 2.} This Lax pair is associated with an underdetermined system.
It is convenient to use a gauge transformation to remove two unknowns from this system initially, we let $c\equiv 1$ and $\ga\equiv 1$. None of the entries of the CC are separated into smaller equations in this case, so the second row of \eqref{cc} gives
\begin{gather*}
\al=a+\dea,\qquad\be=b.
\end{gather*}
Using these values, the equations in the f\/irst row of the CC become
\begin{gather}
\as+\frac{\bs}{b}\de=\bar{a}+\bar{\dea},\qquad\as(a+\dea)-a(\bar{a}+\bar{\dea})+\bs=\bar{b}.\label{l22c2values2}
\end{gather}
There are no further conditions and so there exists one superf\/luous degree of freedom in~\eqref{l22c2values2}. One can use shifts to remove one of the dependent variables, $\de$ say, and write a sole equation in the remaining two dependent variables, $a$ and $b$ say. However, the excess freedom persists, meaning that this system is underdetermined.

{\bf Case 3.} All equations that arise due to the CC in this case are multiplicatively linear and so no genuine nonlinear system can be associated with this Lax pair. However, a false Lax pair is possible in the same way as in case~4, which is considered below. This is because the CC is  the same in both cases~3 and~4 everywhere except for the (1,1) entry, which is split into two equations in case~3, but kept together in case~4. However, the values of $a$ and $\al$, see \eqref{l22c4values2} below, that are found in case 4 are the same as those that would be obtained from the linear equations in case~3. Thus the same false Lax pair is possible for either case.

{\bf Case 4.} This case represents a false Lax pair. 

Splitting the of\/f diagonal entries of~\eqref{cc} as per item~4 in Table~\ref{tab:l22} can be achieved by choosing a Lax pair with the following dependence on the spectral variable
\begin{gather*}
L=\begin{pmatrix} a&\frac{b}{\sqrt{k_1-k_2\nu}}\\\frac{c}{\sqrt{k_1-k_2\nu}}&0\end{pmatrix},\qquad M=\begin{pmatrix}\al&\be\sqrt{k_1-k_2\nu}\\\ga\sqrt{k_1-k_2\nu}&\de\nu\end{pmatrix},
\end{gather*}
where $k_1$ and $k_2$ are constants of proportionality that arise due to splitting the spectral term products in the of\/f diagonal entries. Applying the CC to this brings about the following set of equations
\begin{alignat}{5}
& (1,1)\quad &&\as\al-a\ala+\bs\ga=c\bea,\qquad &&(1,2)_1\quad && k_1\as\be=b\ala,& \nonumber\\
&&& && (1,2)_2\quad && k_2\as\be=\bs \de,& \nonumber\\
& (2,1)_1\quad&& \cs\al=k_1a\gaa,\qquad &&(2,2) \quad && \cs\be=b\gaa,& \nonumber\\
& (2,1)_2\quad&& c\dea=k_2 a\gaa. &&&&&
 \label{l22c4cc}
\end{alignat}

This set of equations can be solved for
\begin{gather}
a=\frac{\ua}{u},\qquad b=\frac{\ua}{v},\qquad c=\frac{\va}{u},\nonumber\\
 \label{l22c4values2}
\al=\frac{k\us}{u}, \qquad\be=\frac{k\us}{k_1 v},\qquad\ga=\frac{k\vs}{k1u},\qquad\de=\frac{\vs}{kv}.
\end{gather}

The unknowns are now $u$ and $v$. As there are no further conditions on the dependent variables, one is free to and choose $v\equiv 1$ and write $a=f(u)$, instead of $a=\bar{u}/u$, where $f(u)$ is an arbitrary function of $u$ and its iterates. Updating the other terms in~\eqref{l22c4values2} accordingly and checking the compatibility condition~\eqref{l22c4cc} shows that any equation, integrable or not, can be artif\/icially associated with this Lax pair by choosing an explicit form of~$f(u)$, which results in the equation $f(u)=\ua/u$. As such, we call it false.

\section{Remaining cases with one zero in each Lax matrix}\label{app:10each}
This appendix deals with the Lax pairs that contain a single zero term in each Lax matrix and yield trivial results.

{\bf B.1.} $L=\begin{pmatrix} aA&bB\\cC&0\end{pmatrix}$, $M=\begin{pmatrix} \al\Al&\be\Be\\\ga\Ga&0\end{pmatrix}$. By setting $D=0$, $\De=0$ in~\eqref{cc}, we see that the CC has simple, linear equations in the (1,2), (2,1) and (2,2) entries. Since each of these entries of the CC contains only two terms, the corresponding pairs of spectral term products must link together, otherwise some other terms must be zero, which leads to a case dealt with by Proposition~\ref{prop00}. In particular we have
\begin{gather*}
B\Ga=C\Be
\end{gather*}
from the (1,2) entry. As such, only two distinct spectral term products appear in the (1,1) entry, $A\Al$ and $B\Ga=C\Be$, each of which is repeated twice. If these do not link together, then all of the equations coming from the CC are linear, for this reason we set
\begin{gather*}
A\Al=B\Ga.
\end{gather*}
The resulting equations arise from the CC \eqref{cc}
\begin{alignat*}{5}
& (1,1)\quad&& \as\al-a\ala+\bs\ga=c\bea, \qquad&& (1,2)\quad&& \as\be=b\ala,& \\
& (2,1) \quad && \cs\al=a\gaa,\qquad &&(2,2) \quad && \cs\be=b\gaa.&
\end{alignat*}
Apart from the (1,1) entry, the equations are linear and, introducing $v(l,m)$ are easily solved for
\begin{gather*}
c=a\va,\qquad\be=\frac{b}{a}\al,\qquad\ga=\al\vs,
\end{gather*}
which cause all three linear equations to become
\begin{gather*}
\as\al=a\ala.
\end{gather*}
This condition means that, despite our best ef\/forts to retain some nonlinearity in the set of equations coming from the CC, the f\/inal equation, from the (1,1) entry, is also multiplicatively linear along with the rest. We conclude that no nonlinear system can arise from this Lax pair.

{\bf B.2.} $L=\begin{pmatrix} aA&bB\\cC&0\end{pmatrix}$, $M=\begin{pmatrix} \al\Al&0\\\ga\Ga&\de\De\end{pmatrix}$. The (2,2) entry of the CC for this Lax pair shows that $B\Ga=0$. Consequently, this case is covered by Proposition~\ref{prop00}.

{\bf B.3.} $L=\begin{pmatrix} aA&bB\\cC&0\end{pmatrix}$, $M=\begin{pmatrix} 0&\be\Be\\\ga\Ga&\de\De\end{pmatrix}$. All equations are linear.

{\bf B.4.} $L=\begin{pmatrix} aA&0\\cC&dD\end{pmatrix}$, $M=\begin{pmatrix} \al\Al&0\\\ga\Ga&\de\De\end{pmatrix}$. Setting $B=\Be=0$ in~\eqref{cc} produces the following set of equations
\begin{alignat*}{5}
& (1,1) \quad && \as\al=a\ala, \qquad& &(1,2) \quad && 0=0, & \\
& (2,1) \quad && \ds \ga D\Ga+\cs\al C\Al=c\dea C\De+a\gaa A\Ga,\qquad &&(2,2) \quad && \ds\de=d\dea,&
\end{alignat*}
where we leave the spectral terms in the (2,1) entry unspecif\/ied at this stage, allowing for this equation to be freely separated into smaller equations later if required. Introducing $u(l,m)$, $v(l,m)$, $\la_i(l)$ and $\mu_i(m)$, the (1,1) and (2,2) entries have the general solution
\begin{gather}
 a=\la_1\va/v,\qquad\al=\mu_1\vs/v,\qquad d=\la_2\bar{u}/u,\qquad\de=\mu_2\hat{u}/u.\label{l12m12}
 \end{gather}
 Substituting \eqref{l12m12} into the remaining equation from the (2,1) entry of the CC, we see that the following quantities arise:
 \begin{gather*}
 \ds\ga=\la_2\frac{\uas\ga}{\us},\qquad\cs\al=\mu_1\frac{\cs\vs}{v},\qquad c\dea=\mu_2\frac{c\uas}{\ua},\qquad a\gaa=\la_1\frac{\gaa\va}{v}.
 \end{gather*}
 If we now introduce
 \begin{gather*}
 x(l,m)=\frac{v\ga}{\us}, \quad\quad y(l,m)=\frac{vc}{\ua},
 \end{gather*}
 then, irrespective of the manner in which we use the spectral terms to separate the equation from the (2,1) entry, only additively linear conditions on~$x$ and $y$ are achievable. Since the (2,1) entry in the f\/inal condition left unsolved, we conclude that this Lax pair can only support linear systems.

\subsection*{Acknowledgements}
The author appreciatively acknowledges support from the Global COE Program ``Education and
Research Hub for Mathematics-for-Industry'' from the Ministry of Education, Culture, Sports, Science
and Technology, Japan.

\pdfbookmark[1]{References}{ref}
\LastPageEnding

\end{document}